# Sensitivity Analysis for Unmeasured Confounding in Medical Product Development and Evaluation Using Real World Evidence

Yixin Fang, AbbVie

Pallavi Mishra-Kalyani, FDA

Xiang Zhang, CSL Behring

Susan Gruber, Putnam Data Sciences

Shu Yang, NCSU

Peng Ding, UC Berkeley

Mingyang Shan, Eli Lilly

Joo-Yeon Lee, FDA

Mark van der Laan, UC Berkeley

Douglas Faries, Eli Lilly

Hana Lee, FDA (The corresponding author; Hana.Lee@fda.hhs.gov)

**Disclaimer:** The views expressed are those of the authors and not necessarily those of the US FDA.

## Abstract

The American Statistical Association Biopharmaceutical Section (ASA BIOP) scientific working group on real-world evidence (RWE) has been making continuous, extended efforts towards a goal of supporting and advancing regulatory science with respect to clinical studies intended to use real-world data for evidence generation for the purpose of medical product development and evaluation (i.e., RWE studies). In 2023, the working group published a manuscript delineating challenges and opportunities in constructing estimands for RWE studies following the framework in ICH E9(R1) guidance on estimand and sensitivity analysis. As a follow-up task, we describe the other issue—sensitivity analysis. Although the FDA's definition of RWE studies includes randomized trials using RWD as a primary source of evidence generation such as pragmatic trials, here we focus on non-randomized RWE studies and the issue of unmeasured confounding which is a major source of bias for most RWE studies. We review the availability and applicability of sensitivity analysis methods for different types of unmeasured confounding. We also provide some practical and regulatory considerations on using sensitivity analysis for such RWE studies. Then, we use a plasmode case study to demonstrate realistic ways to interpret sensitivity analysis results in supporting regulatory decision-making. We conclude with a brief discussion.

## 1. Introduction

In May 2021, the International Council for Harmonisation of Technical Requirements for Pharmaceuticals for Human Use (ICH)—whose mission is to achieve greater regulatory harmonization worldwide to ensure safe, effective, and high-quality medicines development and maintenance—published a guidance on estimand and sensitivity analysis entitled "E9(R1) Statistical Principles for Clinical Trials: Addendum: Estimands and Sensitivity Analysis in Clinical Trials" (ICH 2021). This ICH E9(R1) guidance proposes a unified framework to support the establishment of study objectives, design, conduct, analysis, and interpretation of clinical trials by focusing on two key attributes in clinical trials—estimand and sensitivity analysis. The ICH E9(R1) framework primarily focuses on traditional randomized clinical trials (RCTs).

Since the 21st Century Cures Act and the Prescription Drug User Fee Act (PDUFA) VI, the United States (US) Food and Drug Administration (henceforth, FDA) has been committed to developing a program to evaluate how real-world evidence (RWE), which is defined as *the clinical evidence about the usage and potential benefits or risks of a medical product derived from analysis of real-world data (RWD)* (FDA 2018), can be used to support approval of new indications for approved drugs or to support/satisfy post-approval study requirements. The effort has been continued through PDUFA VII commitment, where the FDA officially launched an *advancing RWE program* to improve the quality and acceptability of RWE-based approaches in support of new intended labeling claims.

Although the principles remain the same, the application of the ICH E9(R1) framework to define estimands and plan/conduct sensitivity analysis in studies designed to generate RWE may not be straightforward. Therefore, the American Statistical Association Biopharmaceutical Section (ASA BIOP) scientific working group on RWE has made a significant effort to discuss a path forward to construct appropriate estimands for RWE studies (Chen et al. 2024). As a follow-up task, the working group has worked on planning and conducting sensitivity analysis for RWE studies.

Here we focus on sensitivity analysis for "non-randomized" RWE studies (henceforth, NR RWE studies), including but not limited to observational studies and single-arm clinical trials with external controls. This is a narrower scope than the FDA's definition of RWE studies which encompasses RCTs using RWD as a primary source to evaluate the safety/efficacy of medical products. For example, the Framework for FDA's RWE Program (FDA 2018) delineates that some (non-traditional) RCTs, such as pragmatic RCTs and registry-based RCTs can also be thought of as RWE studies. Ensuring the validity of the no unmeasured confounder (NUC) assumption has been a key challenge with NR RWE studies. Thus, we would like to have a dedicated discussion on sensitivity analysis to address the impact of unmeasured confounding. All methods discussed here could be applicable when the initial randomization in RCTs is not maintained over the course of study follow-up, and/or there is a systematic difference in study operationalizing characteristics such as treatment adherence and loss-to-follow-up.

The rest of the paper is organized as follows. In Subsection 1.1, we provide a description of sensitivity analysis and specify the focus of this paper—the sensitivity analysis for addressing the impact of the no unmeasured confounding assumption in NR RWE studies. In Subsection 1.2, we present a formal definition of the assumption which could vary by study type. In Section 2, we conduct a literature review of available sensitivity analysis methods while classifying them into three categories depending on the strength of underlying assumptions and the use of external data sources. In Section 3, we discuss practical and regulatory considerations on the choice of appropriate sensitivity analysis methods. In Section 4, we use a plasmode case study to illustrate how to interpret sensitivity analysis results. We conclude the paper with a brief discussion in Section 5.

## 1.1 Definition of sensitivity analysis

ICH E9(R1) defines sensitivity analysis as a series of analyses conducted with the intent to explore the robustness of inferences from the main estimator to deviations from its (1) underlying modelling assumptions and (2) limitations in the data. According to Fang and He (2023), the former issue–deviations from underlying modeling assumptions–could be categorized into deviations from (a) causal modeling assumptions (often known as identifiability assumptions) and (b) statistical modeling assumptions (e.g., when some parametric models are assumed). Furthermore, in the literature of causal inference, causal modeling assumptions often consist of three sets of assumptions: (i) consistency, (ii) positivity, and (iii) NUC (a.k.a., conditional randomization, exchangeability, selection on observables, ignorability of treatment assignment, etc.) Throughout this paper, what we refer to as "the sensitivity analysis" indicates sensitivity analysis aims to examine the impact of deviations from (1) underlying modeling assumptions, particularly deviations from (a) causal modeling assumptions on (iii) the NUC assumption unless otherwise stated. Figure 1 depicts the scope of sensitivity analysis that we intend to cover in this article.

**Figure 1:** The scope of work indicated by bold-faced boxes.

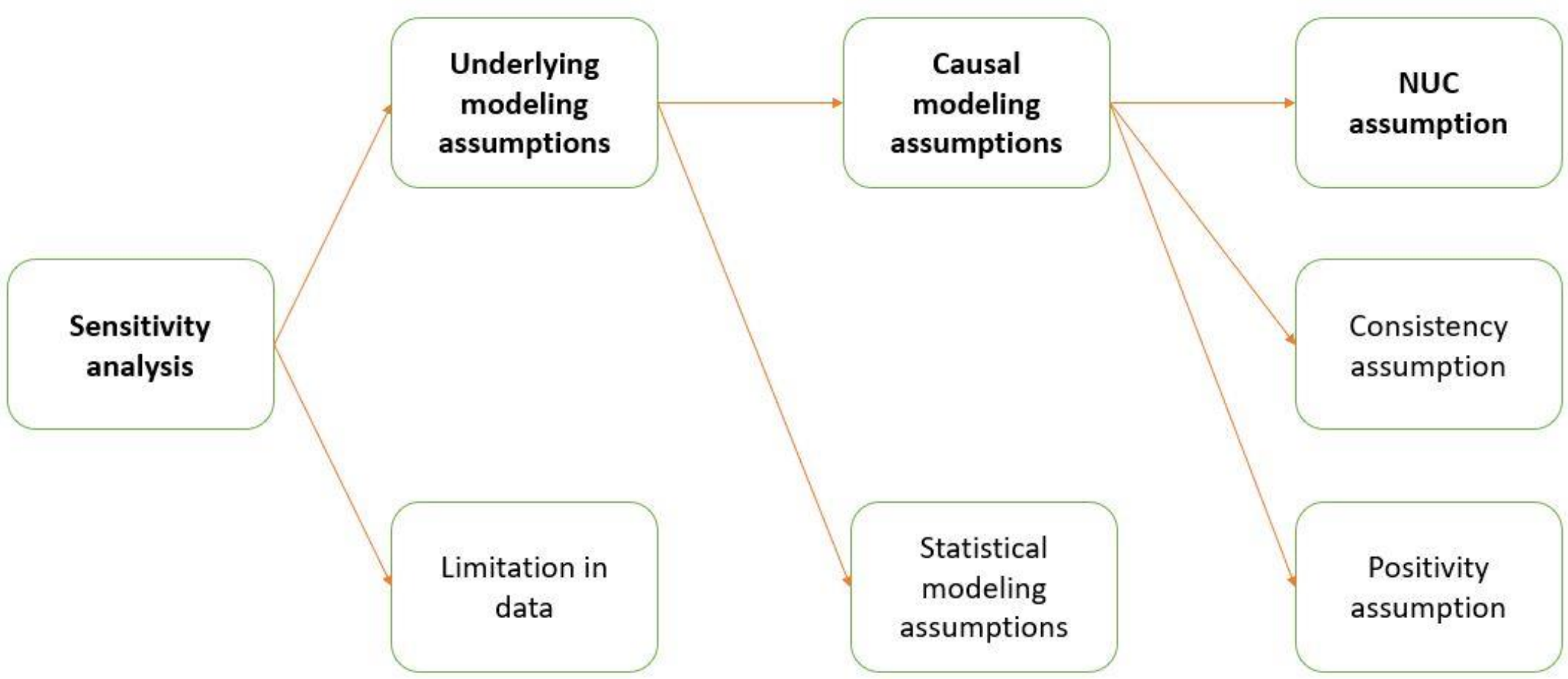

## 1.2 The NUC assumption

Suppose that we are interested in a point-exposure study. Let $X$ be the vector of baseline variables including all the measured confounders. Let $A$ be the treatment variable assumed to be binary for simplicity (1 for investigative treatment and 0 for comparator). Let $Y$ be the outcome variable, which could be of general type (continuous, binary, categorical, ordinal, etc.; see the longitudinal studies for time-to-event outcome). Let $Y^{a=1}$ and $Y^{a=0}$ be the two potential outcomes under a binary treatment strategy of interest, $a = 1$ or 0. In this setting, the NUC assumption can be expressed as:

$$Y^a \perp\!\!\!\perp A|X, \qquad a = 0,1,$$

i.e., the potential outcome $Y^a$ and treatment assignment $A$ are independent given the measured confounders $X$. The NUC assumption states that the measured confounders $X$ capture sufficient information to fully describe the treatment assignment mechanism. Consequently, the treatment ($A = 1$ or 0) can be conceptually viewed as being randomly assigned to study participants, conditional on $X$, as in randomized controlled trials stratified by $X$. In NR RWEs, due to non-randomization, adjusting for measure confounders and evaluating the impact of unmeasured confounders is crucial.

For a longitudinal study, consider visit times $t = 0, 1, \cdots, T$, where $t = 0$ represents the baseline, $t = 1, \cdots, T-1$ represent intermediate visit times to collect treatment status and covariates (which could potentially include intermediate outcome status as well), and $t = T$ be the end of study follow up so that the final outcome status $Y_T$ is measured (but not covariates and treatment status). Let $X_0$ be the vector containing baseline covariates and let $\underline{X_t} = (X_0, \cdots, X_t)$ be the vector consisting of all the observed history up to time $t$ including baseline covariates, time-dependent covariates, and intermediate outcomes. Let $\underline{a} = (a_0, \cdots, a_{T-1})$ be the pre-defined, sequence of the treatment strategy of interest. Let $Y_T^{\underline{a}}$ be the potential outcome at time $T$ for a given treatment sequence $\underline{a}$. Let $\underline{A_t} = (A_0, \cdots, A_t)$ be the vector of actual treatment receipt status up to $t = 0, \cdots, T-1$. Now, the NUC assumption in this setting (a.k.a., sequential exchangeability) can be expressed as,

$$Y_T^{\underline{a}} \perp\!\!\!\perp A_0 | X_0,$$

$$Y_T^{\underline{a}} \perp\!\!\!\perp A_t | \left(\underline{A}_{t-1}, \underline{X}_t\right), t = 1, \cdots, T-1.$$

Similar to the binary treatment setting, the NUC assumption in a longitudinal setting implies that covariate and treatment history information captures sufficient information to determine the next treatment assignment. Therefore, the treatment at each time t (i.e., $A_t$) can considered randomly assigned to study participants, given their history, $\left(\underline{A}_{t-1}, \underline{X}_t\right)$.

Since the NUC assumption is needed in most existing causal inference methods for analyzing data from non-randomized clinical trials (e.g., Phase IV clinical trials, single-arm clinical trials) and real-world studies (e.g., externally controlled clinical trials, and observational studies), conducting sensitivity analysis for the NUC assumption is an important step in generating robust RWE from the analysis of RWD.

## 2. Literature review

Cornfield et al. (1959) conducted the first, formal sensitivity analysis for investigating the causal relationship between cigarette smoking and lung cancer in an observational study. Some early applications of sensitivity analysis in epidemiology are available (Bross 1966; 1967; Schlesselman 1978). Since then, many methods have been proposed, and several comprehensive review papers have been published, including but not limited to Zhang et al. (2018), Zhang et al. (2020), Uddin et al. (2016), Streeter et al. (2017), and D'Agostino McGowan (2022). These review papers provide details of each method, including underlying assumptions, strengths, and limitations. In this paper, we provide another review of the existing methods, by grouping them into three categories, as summarized in Table 1. See below for detailed descriptions.

**Table 1:** Key assumptions, (selected) primary examples, strengths, and limitations for NUC methods under each category

| Category | Assumptions & requirements | Primary examples | Strengths | Limitations |
|---|---|---|---|---|
| Category one | - Parametric methods based on internal data<br>- Parametric modeling of either full or partial relationships among* $A, Y, X,$ and $U$<br>- Distributional assumption on sensitivity parameter for $U$ | - Stratum-specific (Rosenbaum and Rubin, 1983 and others)<br>- Semi-parametric (Dorie et al., 2016)<br>- Simulation-based (Greenland, 2005)<br>- Bayesian (Zhang et al., 2016)<br>- Propensity score-based (Schneeweiss et al., 2009)<br>- Multiple imputation-based (Rubin and Schenker, 1991)<br>- Probabilistic bias analysis (Fox, MacLehose, and Lash 2021) | - Efficient when the parametric and/or distributional assumptions are correct<br>- Do not need external data | - Inadequate if the parametric assumptions are incorrect<br>- Typically, two layers of assumptions are required<br>- Justification is required for both layers of assumptions |
| Category two | - Non-parametric methods based on internal data | - E-value (Ding and VanderWeele, 2016)<br>- G-value (Gruber et al., 2023)<br>- Simple and multidimensional QBA (Fox, MacLehose, and Lash, 2021) | - Robust as no parametric assumption is required<br>- Simple, interpretable, offering visualization w/ or w/o CIs<br>- Do not need external data | - Could be conservative<br>- Some decisions (e.g., threshold selection) require expert knowledge & judgement |
| Category three | - Utilize both internal and (some level of) external data, and,<br>- External data is needed for the purpose of sensitivity analysis | - See Zhang et al. (2018) for review<br>- Methods in Category one fall into Category three if they use external data for sensitivity analysis<br>- Propensity score calibration (Stürmer et al., 2005; 2007) | - Leverage more information, improving effect estimation<br>- Information from external data could serve as a benchmark | - Not applicable if external data to inform sensitivity analysis do not exist<br>- External data should be fit-for-purpose<br>- Requires extra resources and effort |

*Abbreviations: $A$: Treatment; $Y$: Outcome; $X$: Measured covariates; $U$: Unmeasured covariates; CIs: Confidence intervals

## 2.1 Category one: Sensitivity analysis based parametric method using internal data only

Category one includes methods based on sensitivity parameters specifying the relationships among treatment, outcome, measured confounders, and unmeasured confounders. For example, Rosenbaum and Rubin (1983) proposed a method that assumes stratum-specific sensitivity parameters consisting of the prevalence of the binary unmeasured confounder, and its effect on the treatment and the binary outcome, respectively. This method was extended by Imbens (2003) to econometrics. This method was also generalized by Lin, Psaty, and Kronmal (1998) from binary outcomes to other types of outcomes, assuming the true treatment effect can be represented via some generalized linear model that includes the treatment variable as well as the measured and unmeasured confounders. Similar methods have been proposed for other special settings, such as matched observational studies (Rosenbaum 1987). Dorie et al. (2016) considered a semi-parametric version of the method specifying parametric modeling of the unmeasured confounding effects while considering non-parametric relationships among the observed variables. Following the omitted variable bias strategy, Cinelli and Hazlett (2020) developed a sensitivity analysis framework under a linear outcome model and proposed the notion of the robustness value. Methods in this category also include simulation-based sensitivity analysis methods; see Greenland (2005) for an overview.

Some quantitative bias analysis (QBA) methods (Fox, MacLehose, and Lash 2021) that require distributional assumptions on sensitivity parameters (e.g., probabilistic bias analysis) also fall within this category. However, key QBA methods such as simple and multidimensional bias analysis fall under Category two methods, which will be discussed further in the next subsection. Fox, MacLehose, and Lash (2021) extensively reviewed and discussed QBA methods for unmeasured confounding. Discussion on best practices for applying QBA in regulatory settings is available in Lash et al. (2016; 2014).

Roughly speaking, the methods in Category one involve parametric modeling assumptions, in which sensitivity parameters (particularly for unmeasured confounders) are embedded into some distributional forms. Therefore, this category also includes methods based on Bayesian hierarchical regression models (Gustafson et al. 2010; Zhang et al. 2016), propensity score models that incorporate measured variables that are weakly associated with the unmeasured confounders (Schneeweiss et al. 2009; Lee 2014), and implicit multiple-imputation models (Rubin and Schenker 1991). Since the methods in Category one have some parametric components, the methods rely on some implicit "hyper-assumptions", just like the hyperparameters in prior distributions in Bayesian modeling. Therefore, theoretically, we should conduct a second-layer sensitivity analysis to evaluate the robustness of the first-layer sensitivity analysis to the deviation of the hyper-assumptions. The applicability of such methods is therefore insufficient without providing scientific justification (e.g., references or additional guidance) on their use. For this reason, several non-parametric sensitivity analysis methods have been proposed.

## 2.2 Category two: Sensitivity analysis based on non-parametric method using internal data only

The methods in Category two are also based on certain sensitivity parameters, but without parametrically specifying the relationships of unmeasured confounders with treatment and outcome, respectively. Two main representatives of Category two are (1) E-value proposed in Ding and VanderWeele (2016) and VanderWeele and Ding (2017) where 'E' stands for "Evidence" and (2) G-value proposed in Gruber et al. (2023) where 'G' stands for "Gap". Both methods are based on the estimates (point estimate and/or confidence interval [CI] estimate) of the main estimator without depending on how the main estimator is constructed.

E-value is defined as the minimum strength of association that an unmeasured confounder would need to have with both the treatment and the outcome to fully explain away a specific treatment–outcome association, conditional on the measured covariates (VanderWeele and Ding ,  2017). "A large E-value implies that considerable unmeasured confounding would be needed to explain away an effect estimate. A small E-value implies little unmeasured confounding would be needed to explain away an effect estimate." —An excerpt from VanderWeele and Ding (2017).

While E-value is based on the minimum strength of association, G-value is based on the minimal causal gap to alter the causal interpretation of the current findings (Gruber et al. 2023). The causal gap includes potential bias due to violations of key causal assumptions - consistency, NUC, and positivity.  "The G-value is equal to the minimum distance from the null value to either the upper or lower bound of the 95% CI. The G-value can be equivalently expressed in 'causal gap' units on the same scale as the effect estimate in 'SE units' relative to the Standard Error (SE) of the estimate, or, in 'Adj units' relative to the difference between the adjusted and unadjusted estimates." —An excerpt from Gruber et al. (2023).

Another popular approach is simple bias analysis and multidimensional bias analysis methods, which are part of QBA methods (Fox, MacLehose, and Lash 2021). For example, the multidimensional bias analysis provides a bounding factor that could explain away the observed association, under an assumed range of relationships between the exposure (i.e., treatment) and the unmeasured confounding as well as the prevalence of exposed and unexposed (i.e., control).

These methods are simple, interpretable, and offer visual representations with uncertainty estimates such as CIs that facilitate communication among diverse stakeholders. While non-parametric assumption guarantees the robustness of the methods, they can be more conservative than methods in Category one. Also, some judgement may be needed for final decision-making requiring expert discretion (e.g., benchmark selection or plausibility assessment for the assumed range/relationships) or, in some cases, external data such as a literature review. QBA methods might be less preferable for effect measures other than risk ratio (or relative risk; RR) and odds ratio (OR).

Lastly, a method proposed by Chernozhukov et al. (2022) extends the robustness value of Cinelli and Hazlett (2020) to semi-parametric and nonparametric models without assuming a linear form for the outcome.

## 2.3 Category three: Sensitivity analysis utilizing external data

Although the methods in Category one and Category two may refer to external data to establish plausible bounds on the magnitude and direction of bias, external data is not strictly required. Hence, we label all the other methods that require utilizing some external data as Category three; see Zhang et al. (2018) for a comprehensive review of methods in this category. Of note, some methods reviewed by Zhang et al. (2018) can fall into both Category two and Category three depending on the use of external data.

A representative method in Category three is the methods proposed by Stürmer et al. (2005, 2007) that need an additional validation sample to supplement the information on the unmeasured confounding and then calibrate the treatment effect estimator to the validation sample. Since their methods assume a parametric framework, Yang and Ding (2020) provided a more flexible framework. Although not strictly required, Bayesian methods can use external data if the information on unmeasured confounding is captured in the external data (e.g., historical data) which can help inform the prior. However, the external data should also be relevant and reliable, i.e., fit-for-purpose (FDA 2024; 2023c).

Methods in Category three leverage additional information, reducing bias and uncertainty in effect estimation. Information from external data could also serve as a benchmark. However, the existence and validity of external data are warranted to ensure the applicability and reliability of the methods. Also, these methods typically require extra resources and efforts to utilize.

## 2.4 Sensitivity analysis for longitudinal studies

We are now ready to review sensitivity analysis methods for the NUC assumption for longitudinal studies (a.k.a. sequential exchangeability). In this setting, the following two types of treatment regimes are often of interest—static treatment regimes (STRs) and dynamic treatment regimes (DTRs). Tsiatis et al. (2019) provided a comprehensive review of existing methods for estimating the effect of DTRs. However, the literature for conducting sensitivity analysis to evaluate the robustness of the effect of DRTs under the violation of the sequential exchangeability assumption is lacking, with few exceptions, e.g., Kallus and Zhou (2020).

Therefore, we focus on sensitivity analysis methods for the NUC assumption under the objective of comparing two STRs, say $\underline{a}^1 = \underline{1}$ vs. $\underline{a}^0 = \underline{0}$. The two main challenges in this setting are handling (a) intercurrent events and (b) confounding bias. For handling of intercurrent events, ICH E9(R1) proposed five strategies and Fang and He (2023) reviewed sensitivity analysis methods for the underlying assumptions made behind these strategies. For confounding bias adjustment, efficient methods such as g-methods (Hernan and Robins 2020) and the targeted maximum likelihood estimator (TMLE; van der Laan and Rose 2011) are applicable. Some of these methods are parametric and can be considered as extensions of methods in Category one. For example, if a marginal structural model (Robins 1999) via inverse probability treatment weighting (IPTW, or augmented IPTW) are applied to construct the main estimator, then we can conduct the sensitivity analysis using existing methods such as Robins et al. (2000) and Brumback et al. (2004). If a structural nested model (Robins 1994) is assumed and the g-estimation method (Robins 1998a) is applied, then we can conduct the sensitivity analysis using Robins (1998b) and Yang and Lok (2018).

Despite the existence of a research gap, we believe that most methods in Category two and Category three that were originally proposed for point-exposure studies can be extended to longitudinal studies. For example, if the g-formula is applied to the formulation of an estimand (Hernan and Robins 2020; van der Laan and Rose 2011), we can conduct the sensitivity analysis using existing methods such as Díaz et al. (2018). We also look forward to the development of the E-value for the NUC assumption in longitudinal studies.

## 3. Practical and regulatory considerations for addressing unmeasured confounding in the NR RWE studies

One of the key goals in the formulation of NR RWE studies is to prevent sources of bias as much as possible at the design stage. Design considerations include the selection of best-quality data sources that ideally capture information on all relevant confounders so that any observed difference from an appropriate statistical analysis can be interpreted as the (causal) effect of the medical product. However, in practice, it is often implausible to fully characterize all confounding variables upfront and/or to identify a perfect data source that captures all those confounding variables at the design stage. Even with sufficient knowledge of confounding variables and the presence of the best data source, the (known) confounding variables could be inconsistently measured (measurement error or missingness) or not measured at all in the data.

With these real-life challenges in mind, we address practical and regulatory considerations for assessing the impact of violating the NUC assumption by (1) first outlining different categories of unmeasured confounding variables, followed by (2) delineating key elements for implementing a sensitivity analysis plan for NR RWE studies in regulatory submissions.

## 3.1 Potential sources of unmeasured confounders

Schneeweiss (2006) categorized unmeasured confounders into two categories: (a) unmeasured in the current study but measurable in a validation study and (b) unmeasured or unmeasurable. Zhang et al. (2020) further refined these into three possible categories regarding available information on unmeasured confounders: (i) researchers do not know any information about the unmeasured confounders, (ii) the data does not have collected information for confounding variables for every study participant but does for a subsample, and (iii) the data does not have collected information for the confounding variable for any patient, but such information is available in data external to the study database.

Motivated by Schneeweiss (2006) and Zhang et al. (2020), we provide a finer categorization of unmeasured confounders: (1) unknown confounders, (2) known to some extent but unmeasured/unmeasurable confounders, (3) known but unmeasured/unmeasurable confounders with proxies in the current dataset, (4) known confounders, that are not measured in the current dataset, but some (either patient-level data or summary data) information regarding the relationships of these confounders is available from an external database(s), and (5) confounders which are not measured for all patients but only measured for a subsample of the patients in the current dataset.

As mentioned earlier, Cornfield et al. (1959) conducted the first sensitivity analyses to evaluate the impact of unmeasured confounding on the estimated RR of lung cancer between smokers and non-smokers. We use this example and consider hypothetical scenarios to describe how each of the above categorizations of unmeasured confounders could be operationalized in practice.

### Scenario 1:

In this scenario, we suspect there could be unmeasured confounders such as genetic, environmental, or societal factors, but we don't exactly know what they are due to scientific gaps in existing research. Many methods reviewed earlier can be applied to this scenario. In particular, any tipping point analyses may be appropriate, e.g., the array approach (Schneeweiss 2006) and E-value (VanderWeele and Ding 2017). The array approach produces a grid of estimates based on a given range of underlying parameters that measure the relationship among unmeasured confounders, smoking, and lung cancer. As described in Section 2.2, the E-value assesses the strength of unmeasured confounding that could explain away the observed relative risk (after adjusting for measured confounders) between smoking and lung cancer. These methods may require benchmark values to assist in interpretating the results. The benchmark could be tied to the most influential measured confounder, so one could use its relative risk obtained from univariate analysis with the outcome to calculate the specific benchmark value. To be conservative, researchers often take a large benchmark or a wide range of underlying parameters when applying the tipping point analyses.

### Scenario 2:

In this scenario, there is a known confounder, but the factor is either unmeasurable or unavailable in the current data sources. For example, Fisher (1958) suspected that some genotypes might be unmeasured confounders of the relationship between lung cancer and smoking. In this scenario, some information is available regarding the relationship between the unmeasured confounder and outcome, e.g., genetic literature suggests that the genetic effect on lung cancer would not be significantly larger than other biomarkers. Therefore, we can apply the same tipping point analyses as before but potentially based on a narrower range to apply the array approach or a smaller benchmark value to interpret the E-value.

### Scenario 3:

In this scenario, the current database lacks information on a known confounder, but includes information on a proxy. The degree of associations among the proxy, exposure, and outcome can inform the range of underlying parameters in sensitivity analyses. For example, suppose that the "smoking-gene" is the unmeasured confounder, and the factor is known to be highly associated with injection drug use (IDU; the proxy). Then, we can estimate associations between IDU and smoking, IDU and lung cancer, respectively. These estimates can then serve as a numerical basis to establish an appropriate range for the array approach and a benchmark for the E-value.

### Scenario 4:

In this scenario, the current database lacks information on a known confounder, but an external database provides it. Like in scenario 3, we can use information from the external data to estimate and provide a numerical basis to propose an appropriate range or benchmark value. For example, Fisher (1958) described a small twin study that estimated the association between the smoking-gene and smoking. For more recent application, see Zhang et al. (2016) that used Bayesian methods to produce an adjusted treatment effect estimate.

### Scenario 5:

In this scenario, the suspected confounder information is not available from all participants in the current database but is available from a subsample of the participants (i.e., internally validated data). Like scenario 3 and 4, we can estimate the relationships among the confounder, exposure, and outcome using the subsample to inform a range or a benchmark value in sensitivity analysis for the whole sample. For example, see Faries et al. (2013) for the use of a subsample of data with complete biomarker (HbA1c) information to compare costs of care between two treatments for type 2 diabetes. Beyond benchmarking, there is a selection of methods that could produce adjusted effect estimates when internally validated information is available (e.g., Bayesian approaches, multiple imputation, and TMLE), assuming there is no systematic difference in collecting the confounder data between the subsample and the rest.

### 3.2 Practical and regulatory considerations for choosing sensitivity analysis

In this section, we discuss practical and regulatory considerations for choosing appropriate sensitivity analysis methods for the NUC assumption in the NR RWE studies.

#### Interpretability and Communication:

As all sensitivity analyses for the NUC assumption are based on untestable assumptions, including assumptions with respect to sensitivity parameters, the interpretation of results produced by the analyses is contingent upon inherent subjectivity. Therefore, it may be preferable to consider methods that rely on fewer assumptions or sensitivity parameters.

The sensitivity analysis requires input from clinicians or epidemiologists regarding the clinical relevance or plausibility of the underlying assumptions, such as a reasonable range of sensitivity parameters (e.g., a range for the array approach and a benchmark value for the E-value). Therefore, the assumptions and sensitivity parameters should be interpretable for these professionals, and the results of the analysis should be conveyed easily. This can facilitate their understanding of the results and thus impact regulatory decision-making.

It is also important to provide results of the sensitivity analysis, including any limitations or caveats, in a reasonably comprehensible language/format to ensure that the results are understandable to relevant stakeholders, including healthcare providers and patients. Therefore, methods that could provide a summary and visual representation might be preferable.

#### Applicability and Implementation:

Another crucial factor to consider is applicability and flexibility, as well as ease of implementation.

Certain sensitivity methods can work with a variety of summary measures, such as RR, OR, or hazard ratio (HR), while others have limited applicability. Similarly, some methods are only applicable to certain types of confounders (e.g., binary). Some methods are only suitable for a single unmeasured confounder (e.g., array approach in Schneeweiss 2006), while others could handle multiple confounders (e.g., E-value, G-value). The choice of a sensitivity method should consider alignment with the summary measure in the study estimand, and thus with the main estimator for the final (outcome) analysis. If multiple unmeasured confounders are considered, it might be preferable to consider a method that could handle various types of variables, such as binary, categorical, or continuous variables.

Finally, the availability of statistical software or programming may facilitate the implementation of the selected method. This is also important from a reproducibility perspective, which is one of the key considerations in regulatory settings.

### Data fit-for-purpose and transparency:

The FDA's guidance documents on RWD (FDA 2024; 2023c) stress the importance of the reliability and relevance of the data (i.e., data fit-for-purpose). Considerations regarding the data fit-for-purpose apply not only to the data used primary analysis but also to any analysis used to inform the strength and quality of the evidence, including sensitivity analysis. Therefore, the data used for sensitivity analysis needs to be relevant and reliable. If external data sources are considered for the sensitivity analysis (i.e., methods in Category three), the relevance and reliability of the external data should be carefully examined.

In certain cases, data might only be available at a summary level for patient privacy purposes. As some methods are only applicable to patient-level data, the choice of a sensitivity method should consider the level of availability (summary vs patient-level) in the selected data. For example, the PS calibration method (Stürmer et al. 2005) requires patient-level data from an external source(s), whereas the E-value method (VanderWeele and Ding 2017) only requires the observed treatment effect estimates (e.g., RR, OR, and HR).

Finally, the FDA guidance documents on regulatory considerations for the use of RWD/RWE (FDA 2023b) as well as external control (FDA 2023a) emphasize submission of patient-level data for any RWD analyzed as part of the clinical study included in a marketing application, as required under FDA 508 regulations (21 CFR 314.50(f) and 601.2). Therefore, it may be helpful to consider the upfront submission of all necessary RWD for FDA review. For example, if certain RWD used for the sensitivity analysis are owned and controlled by third parties, it is crucial for sponsors to establish agreements with those parties to ensure that all relevant data can be provided to the FDA.

## 4. Interpretation of the sensitivity analysis results

This section will not delve into the comparative performance of the sensitivity analysis methods reviewed in Section 2. Instead, it will focus on the interpretation of the results obtained from a specified sensitivity analysis method. A key aspect of the discussion is the pre-specification of the primary estimation method and sensitivity analysis method in the statistical analysis plan (SAP) before the statistician or analyst examines the outcome data. For illustration, we consider an observational study aimed at estimating the average treatment effect (ATE) for regulatory purposes. We assume that the data is fit-for-purpose. We also assume that TMLE (van der Laan and Rose 2011) was specified as the primary analysis method, and E-value (VanderWeele and Ding 2017) along with G-value (Gruber et al. 2023) were both specified as the sensitivity analysis methods for the NUC assumption in the study SAP. This illustration bypasses the implementation details of TMLE, including the specification of a super-learner library and instead focuses on determining thresholds for E-value interpretation.

## 4.1 A plasmode case-study

In another paper contributed by the ASA RWE working group, Ho et al. (2024) generated a plasmode dataset to illustrate the causal inference roadmap, where the final step is for interpretation of the sensitivity analysis. In Ho et al. (2024), G-value (Gruber et al. 2023) was applied to conduct the sensitivity analysis without providing any benchmark value for the interpretation. In this section, we reuse their plasmode example to further elaborate the sensitivity analysis results.

Ho et al. (2024) generated the plasmode dataset using data from the START trial (NCT00409188), which a multicenter, randomized, double-blind placebo-controlled study of the cancer vaccine Stimuvax (L-BLP25 or BLP25 Liposome Vaccine) in nonsmall cell lung cancer subjects with unresectable stage III disease. The START trial data is available on *Project Data Sphere* (https://www.projectdatasphere.org/), an open-access platform that aggregates trial data and provides free access to the research community. For most trials, Project Data Sphere provides only the control arm datasets, including the START trial. Therefore, Ho et al. (2024) used the START's control arm data as a "blueprint" to simulate an observational cohort study incorporating the real control arm and the simulated treatment arm. They introduced a certain level of confounding bias based on a set of measured confounders.

In their plasmode example, the investigational treatment is the cancer vaccine Stimuvax ($A = 1$) and the control treatment is assumed to be the standard of care ($A = 0$), whereas the START was a placebo-controlled trial. The simulated data includes $n_1 = 1000$ subjects in the treated arm and $n_0 = 500$ in the control arm. As described in Ho et al. (2024), the set of nine measured confounders (X) includes: Age (in years), sex (0 for female and 1 for male), smoking (0 for no and 1 for yes), histology (0 for other and 1 for squamous cell carcinoma), ECOG performance status (0 for full active and 1 for restricted), advanced stage (0 for stage IIIA and 1 for stage IIIB or IIIC), type of chemo (0 for concomitant and 1 for sequential), chemo response (0 for no and 1 for yes), and census region (0 for rest of world, 1 for West Europe, and 2 for north America).

## 4.2 Estimand, estimator, and sensitivity analysis

For illustrative purposes, suppose that the primary endpoint is survival status at 60 months from treatment initiation, with $Y = 1$ for survival and $Y = 0$ for death. Let $Y^a$ be the potential outcome under treatment $a = 1$ or $0$. Suppose that the estimand, say $\theta$, is the RR of survival at 60 months when all patients in the study were adhered to their initial treatment assignment and retained by the end of the study follow-up:

$$\theta = P(Y^1 = 1)/P(Y^0 = 1).$$

With this estimand, estimand attributes are as follows: (1) Population is the nonsmall cell lung cancer patients with unresectable stage III disease, more specifically defined via a set of inclusion/exclusion criteria in the START trial; (2) treatment variable is $A$ (0 or 1); (3) outcome variable is the survival status at 60 months; (4) intercurrent events (e.g., treatment switching, treatment discontinuation) are handled by the hypothetical strategy; and (5) the population-level summary is RR. As mentioned earlier in this section, TMLE is the primary method of estimation for $\theta$; E-value and G-value are the sensitivity analysis method for the NUC assumption.

Of note, $P(Y = 1|A = 1)$ and $P(Y = 1|A = 0)$ can be estimated via the Kaplan-Meier (KM) curves, providing an unadjusted estimator (say, $\hat{\theta}_{un}$) of the RR between $Y$ and $A$. However, this unadjusted estimator may be biased estimate of $\theta$ in the presence of confounding, which is true in this plasmode example. Although the unadjusted approach is not part of the pre-specified method, we describe it here for later discussion in Section 4.3 – strategies for planning and interpreting the sensitivity analysis.

Now suppose that we implement the Kaplan-Meier estimator and TMLE and obtain point estimates $\hat{\theta}_{un} = 1.87$ with 95% CI = $(1.64, 2.13)$ and $\hat{\theta}_{TMLE} = 1.68$ with 95% CI = (1.55, 1.82), respectively. Although directions of effects are consistent (i.e., treatment results in longer survival and thus beneficial), difference in size of estimates may indicate potential bias.

Now let's focus on the sensitivity analysis using $\hat{\theta}_{TMLE}$ as the primary effect estimate. Formula for the E-value for RR is given by

$$E = RR + \sqrt{RR \times (RR - 1)}.$$

Accordingly, the E-value associated with our primary effect estimate $\hat{\theta}_{TMLE}$ = 1.68 is $1.68 + \sqrt{1.68(1.68 - 1)} = 2.75$. This means that the observed survival RR of 1.68 could be explained away by an unmeasured confounder that was associated with both the treatment and the outcome by an RR of 2.75-fold each. The decision on whether such an unmeasured confounder exists and whether the RR of 2.75-fold is considered weak or strong requires expert judgement.

The E-value can also be examined for the lower bound of the 95% CI (of RR) = 1.55. The formula is the same, and therefore $1.55 + \sqrt{1.55(1.55 - 1)} = 2.47$ is the E-value for the lower bound. It indicates that that an unmeasured confounder would need to have a RR of at least 2.47 (with both the treatment and outcome) to shift the result toward no effect.

See R package "EValue" or website https://www.evalue-calculator.com/ developed by VanderWeele and Ding (2017) and Mathur et al. (2018) for more instruction on the implementation of E-value.

Next we look into G-value results shown in Figure 2, which depict the change in the effect estimate and 95% CI under different levels of (assumed) causal bias. As we consider the hypothetical strategy (i.e., under perfect treatment adherence) and the plasmode dataset satisfies the positivity assumption (i.e., P(Y|A, X)>0 for all A=a, X=x), the causal bias in this context represents bias due to unmeasured confounding. Figure 2 illustrates the change in the effect estimate $\hat{\theta}_{TMLE} = 1.68$, and 95% CI under different assumed causal biases. The causal gap is shown on the x-axis in RR units ($\delta$), and relative to the bias reduction when adjusting for measured confounders (*Adj units)*.

As depicted in Figure 2, the G-value of 0.55 with 2.89 Adj units. This implies that residual bias due to unmeasured confounder(s) should be larger than 0.55 in RR unit, (or 2.89 times stronger than the bias adjusted for by measured confounders) to negate the observed finding. Similar to the E-value, assessing the plausibility of the causal gap, such as the existence and strength of unmeasured confounder(s), requires expert judgement.

The G-value is calculated by finding the minimum distance from the null to the upper and lower bounds on the confidence interval. Its sign is negative if the closest bound is smaller than the null value, and positive if the closest bound is greater than the null value. The G-value calculation will be available in the *tmle* package on CRAN beginning with v 2.0.2.

Figure 2. Targeted learning-based causal gap analysis

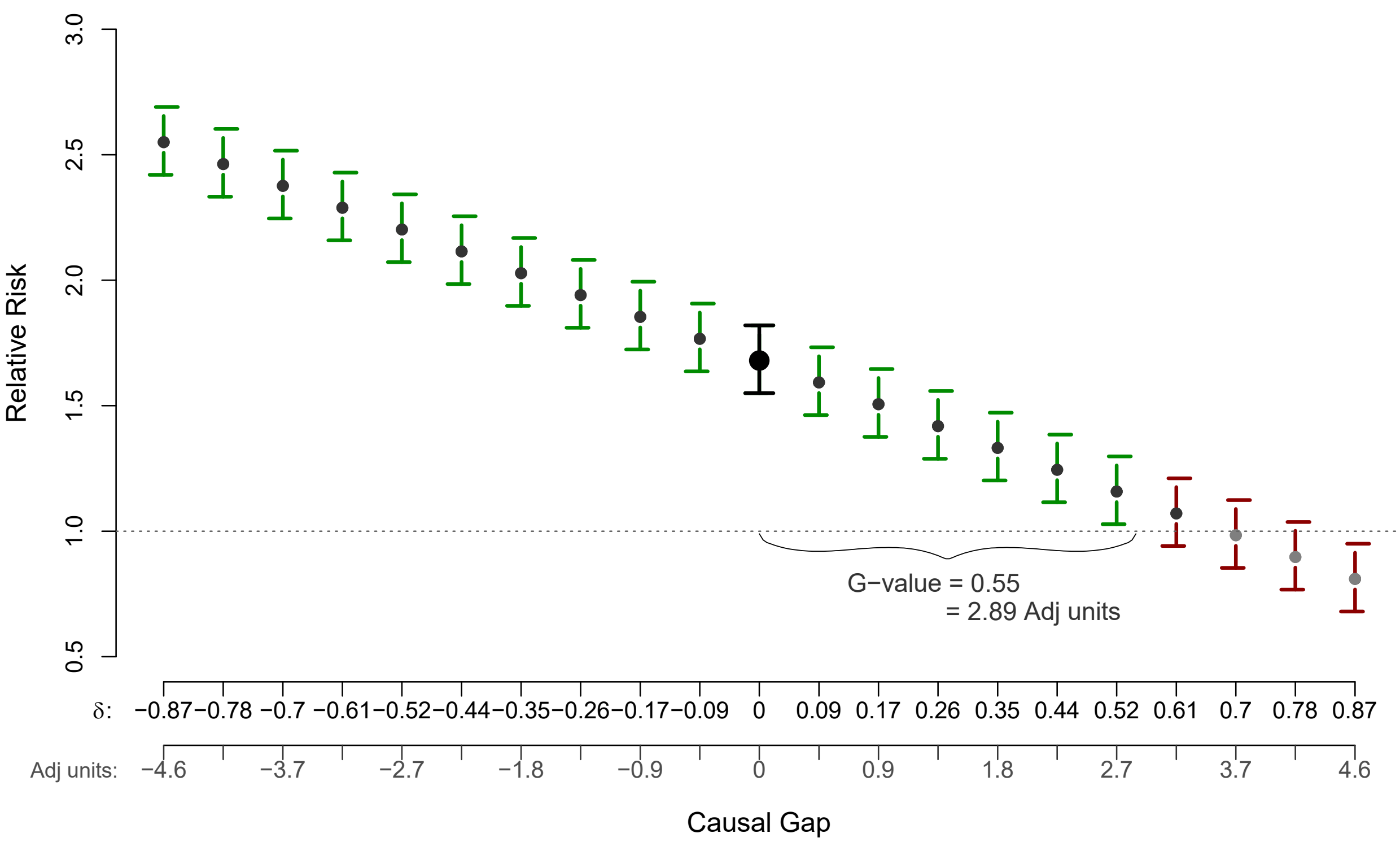

The key challenge with the use of E-value or G-value lies in determining the existence and magnitude of E- or G-value necessary to assert or negate study findings. Pre-specification of a benchmark value to determine the robustness of findings is challenging. In the next subsection, we offer several strategies to pre-specify a benchmark for the E-value and G-value in the SAP.

### 4.3 Some strategies for interpreting the sensitivity analysis results

In this subsection, we discuss four strategies for planning and interpreting the sensitivity analysis results:

(1) Specifying a predetermined threshold supported by a literature review
(2) Specifying a fixed threshold supported by historical results
(3) Specifying a method for identifying a threshold based on internal data
(4) Specifying a method for identifying a threshold based on external data

In terms of efforts needed, the order is (1) < (2) < (3) < (4). In terms of reliability, the order is also (1) < (2) < (3) < (4). Regarding data availability, (4) is the most challenging to apply because it relies on external data. Since the threshold is pre-specified by (1) and (2) before the study is conducted, the first two strategies could be more reliable than the other two. For (3) and (4), as emphasized in Section 3.2, internal data and external data should be fit-for-purpose.

#### Specifying a predetermined threshold supported by literature review

In the SAP, we may specify a fixed threshold determined by the discussion between the sponsor and the regulatory agency based on the literature review provided by the sponsor. For example, we may predetermine the E-value threshold, say $\Delta$, as 2 in the SAP supported by the literature review showing that the largest effect in terms of RR that a risk factor could have on the outcome variable was at most 2. If the resulting E-value is larger than $\Delta$ (e.g., in the above example, the E-value associated with CI's lower level is 2.47, which is larger than 2), we can claim that the results of the main estimator are robust. A similar argument applies to the G-value, where bounds on the effect size can be established based on relevant prior studies in a related population or related disease, effect sizes in single-arm trials, etc.

#### Specifying a fixed threshold supported by historical results

In the SAP, we may specify a fixed threshold supported by preliminary results or the meta-analysis of similar studies provided in the literature, like the process of specifying a non-inferiority margin. For example, using a cohort study reviewed in the literature with similar outcome variables and treatments, we can calculate the ratio of the unadjusted RR estimate $RR_{un}$ (e.g., RR based on Kaplan-Meier estimates of $P\{Y=1|A=1\}$ and $P\{Y=1|A=0\}$ in Section 4.2) and the adjusted RR estimate $RR_{adj}$ to estimate the bounding factor $B = \max\left\{\frac{RR_{un}}{RR_{adj}}, \frac{RR_{adj}}{RR_{un}}\right\}$, which accounts for the bias due to the measured confounders. Then we use

B to calculate a threshold for the E-value, $\Delta = B + \sqrt{B(B-1)}$. As a hypothetical example, suppose that the estimated $B$ from a similar study in the literature is 1.25, which leads to $\Delta = 1.25 + \sqrt{1.25(1.25-1)} = 1.81$. We may then specify $\Delta = 1.81$ in the SAP, providing the reference and the reasoning as scientific justification for the chosen threshold. If the resulting E-value is larger than $\Delta$ (e.g., in the above example the E-value associated with CI's lower level is 2.47, which is larger than 1.81), we can claim that the results of the main estimator are robust. Comparing the unadjusted estimator with the adjusted estimator is a conservative strategy to specify the threshold because the E-value is the threshold for the residual unmeasured confounding given the measured confounders. Alternatively, we can compute the threshold based on the leave-one-out strategy; see Chapter 17 of Ding (2024).
A similar argument applies to the G-value, where bounds on the effect size can be established based on knowledge of the natural history of the disease, such as spontaneous remission rates.

### Specifying a method for identifying a threshold based on internal data

In the SAP, we may specify a method for identifying a threshold using the results obtained from the current study. For example, we may specify the following procedure:

(1) Obtain the unadjusted estimate $\hat{\theta}_{un}$ and adjusted estimate $\hat{\theta}_{TMLE}$

(2) Obtain the ratio of them, denoted as $\hat{B} = \max\left\{\frac{\hat{\theta}_{un}}{\hat{\theta}_{TMLE}}, \frac{\hat{\theta}_{TMLE}}{\hat{\theta}_{un}}\right\}$

(3) Use $\hat{B}$ to find a threshold, $\hat{\Delta} = \hat{B} + \sqrt{\hat{B}(\hat{B}-1)}$

Continued with the above plasmode example, if the current study provides $\hat{\theta}_{un} = 1.87$ and $\hat{\theta}_{TMLE} = 1.68$, we can calculate that $\hat{B} = 1.11$ and $\hat{\Delta} = 1.46$. Since the E-value associated with the lower 95% CI is 2.47, which is larger than 1.46, we can claim that the results of the main estimator are robust.

For G-value, internal data can be used to investigate the impact of unmeasured confounding on the effect estimate scale ($\delta$) by re-analyzing the data after dropping several known measured confounders and noting whether the point estimate shifts toward or away from the null, and by how much.

### Specifying a method for identifying a threshold based on external data

In the SAP, we may specify a method for identifying a threshold using external data. As a hypothetical example, suppose there are partially available external data on some unmeasured confounders, which are believed to be potential confounders but unmeasured in the current study database. Then we may specify the following procedure:

(1) Obtain a new estimate, $\hat{\theta}_{new}$ using a combined (internal plus external) data

(2) Obtain the ratio of the new estimate and TMLE, denoted as $\hat{B} = \max\left\{\frac{\hat{\theta}_{new}}{\hat{\theta}_{TMLE}}, \frac{\hat{\theta}_{TMLE}}{\hat{\theta}_{new}}\right\}$

(3) Use $\hat{B}$ to find a threshold, $\hat{\Delta} = \hat{B} + \sqrt{\hat{B}(\hat{B}-1)}$

Continued with the plasmode example, if combining the current study and the external data provides $\hat{\theta}_{new} = 1.43$ and $\hat{\theta}_{TMLE} = 1.68$, it leads to $\hat{B} = 1.17$ and $\hat{\Delta} = 1.62$. Since the E-value for the lower bound of 95% CI is 2.47, which is larger than 1.62, we can claim that the results of the main estimator are robust. The difference $\hat{\theta}_{new} - \hat{\theta}_{TMLE} = -0.25$ is less than the G-value of 0.55, which increases confidence that the results of the main estimator are robust.

## 5. Discussion

NR RWE studies for medical development and evaluation should incorporate quantitative assessment of potential bias due to unmeasured confounding. This paper aims to provide a reasonable categorization of the existing methods, clarify operational characteristics of different methods, and elucidate principles for differentiating and defining the underlying traits of potential unmeasured confounders.

We also delineate key practical and regulatory considerations to guide the selection of sensitivity analysis methods for NR RWE studies – (1) interpretability and communication, (2) applicability and implementation, and (3) data fit-for-purpose and transparency. These are high-level considerations and recommendations. While such generality ensures that our discussions are relevant across various contexts, we acknowledge that it may lack specific details. Given that regulatory considerations can vary by context, early engagement with the regulatory agency can be invaluable in planning and conducting sensitivity analyses.

Although our illustration in Section 4 primarily focuses on the E-value and G-value, other methods are also applicable. For implementation of QBA methods, see Kawabata et al. (2023) that provides a systematic review of QBA software development between 2011 and 2021. To our knowledge, the other methods mentioned in this manuscript may not be readily implementable using existing software.

We note that there are other causal inference methods that do not require the NUC assumption, e.g., the instrumental-variable methods (Imbens and Angrist 1994; Imbens 2014), the front-door-criterion (Pearl 2000), interrupted time-series analysis (Cook, Campbell, and Day 1979; Gillings, Makuc, and Siegel 1981; Wagner et al. 2002), and negative controls (Rosenbaum 1989; Groenwold 2013; Lipsitch, Tchetgen, and Cohen 2010; Prasad and Jena 2013; Schuemie et al. 2014; Tchetgen Tchetgen et al. 2020). However, these methods are beyond the scope of this paper, because they require another corresponding set of identifiability assumptions other than the NUC assumption.